\documentclass[sigconf]{acmart}

\AtBeginDocument{%
  }

\setcopyright{acmlicensed}
\copyrightyear{2018}
\acmYear{2018}
\acmDOI{XXXXXXX.XXXXXXX}
\acmConference[Preprint]{Preprint}{2026}{arXiv}
\acmISBN{978-1-4503-XXXX-X/2018/06}

\usepackage{placeins}

\usepackage{xcolor} 

\usepackage{subcaption}

\usepackage[most]{tcolorbox}
\usepackage{listings}
\lstdefinestyle{prompt}{
  basicstyle=\ttfamily\footnotesize,
  breaklines=true, columns=fullflexible, keepspaces=true,
}

\definecolor{jbKw}{HTML}{0033B3}   
\definecolor{jbFn}{HTML}{00627A}   
\definecolor{jbId}{HTML}{871094}   
\definecolor{jbStr}{HTML}{067D17}  
\definecolor{jbCom}{HTML}{8C8C8C}  
\definecolor{codebg}{HTML}{FFFFFF} 
\definecolor{coderule}{HTML}{C8C8C8}

\lstdefinestyle{sqlpretty}{
  language=SQL,
  alsoletter={_},
  backgroundcolor=\color{codebg},
  basicstyle=\ttfamily\scriptsize\color{black},
  keywordstyle=\color{jbKw}\bfseries,
  morekeywords=[2]{FORMAT_DATE,PARSE_DATE,ANY_VALUE,CURRENT_TIMESTAMP,SUBSTR,CAST,COUNT,LOWER,TIMESTAMP},
  keywordstyle=[2]\color{jbFn}\itshape,
  morekeywords=[3]{DocumentIdentifier,SourceCommonName,TranslationInfo,V2Locations,V2Themes,V2Organizations,_PARTITIONTIME,date,url,domain,mentions,art},
  keywordstyle=[3]\color{jbId},
  stringstyle=\color{jbStr},
  commentstyle=\color{jbCom}\itshape,
  morecomment=[l]{--},
  numbers=none,
  frame=single, framesep=5pt, rulecolor=\color{coderule},
  showstringspaces=false,
  breaklines=true, breakatwhitespace=true,
  columns=fullflexible, keepspaces=true,
  xleftmargin=3pt, xrightmargin=3pt
}

\lstdefinestyle{promptpretty}{
  backgroundcolor=\color{codebg},
  basicstyle=\ttfamily\scriptsize\color{black},
  numbers=none,
  frame=single, framesep=5pt, rulecolor=\color{coderule},
  showstringspaces=false,
  breaklines=true, breakatwhitespace=true,
  columns=fullflexible, keepspaces=true,
  xleftmargin=3pt, xrightmargin=3pt
}

\tcbuselibrary{listings}   
\newtcblisting{sepprompt}{
  listing engine=listings, listing only,
  listing options={
    basicstyle=\ttfamily\scriptsize,
    breaklines=true, breakatwhitespace=true,
    columns=fullflexible, keepspaces=true,
    numbers=none, showstringspaces=false
  },
  colback=green!8, colframe=black!80, boxrule=0.6pt, arc=2pt,
  left=5pt, right=5pt, top=3pt, bottom=3pt,
  title={SEP System Prompt (LLM)}, fonttitle=\bfseries\small,
  coltitle=white, colbacktitle=black!85, breakable
}

\renewcommand\footnotetextcopyrightpermission[1]{}

\begin{document}

\title{Investigating Forecasting Capabilities of Narrative Networks on Emerging Stories}





\author{Ravi Teja Karumuri}
\affiliation{%
  \institution{Arizona State University}
  \department{School of Computing and Augmented Intelligence}
  \city{Tempe}
  \state{Arizona}
  \country{USA}}
\email{rkarumur@asu.edu}

\author{Steven Corman}
\affiliation{%
  \institution{Arizona State University}
  \department{Hugh Downs School of Human Communication}
  \city{Tempe}
  \state{Arizona}
  \country{USA}}
\email{Steve.Corman@asu.edu}

\author{Huan Liu}
\affiliation{%
  \institution{Arizona State University}
  \department{School of Computing and Augmented Intelligence}
  \city{Tempe}
  \state{Arizona}
  \country{USA}}
\email{huanliu@asu.edu}

\author{Joshua Garland}
\affiliation{%
  \institution{Arizona State University}
  \department{Center on Information and Narrative Complexity}
  \city{Tempe}
  \state{Arizona}
  \country{USA}}
\email{Joshua.Garland@asu.edu}


\renewcommand{\shortauthors}{Karumuri et al.}


\begin{abstract}
Networks have been used to model the interactions among the elements that underlie a narrative. In this paper, we investigate whether the structure of these networks can predict which links will emerge as a storyline evolves. We construct these networks from news articles using a Semantic Edge Parser (SEP) built on a Large Language Model (LLM), and evaluate the networks on three disparate real-world storylines, the 2022 Canadian Freedom Convoy, Hurricane Melissa (2025), and the 2025–26 Iranian protests. To test the narrative network’s predictive capability, we construct a rolling window of the narrative and predict the emergent links between existing narrative elements on the day following the rolling window period. We identify candidate links with seven standard link predictors, including the Adamic-Adar index, Common Neighbors, and Preferential Attachment. Forecasting next-day links across the three storylines, we find Preferential Attachment to be the strongest standard link predictor, achieving a mean ROC-AUC of 0.76–0.79. It exceeds EdgeBank, a memorization baseline, by 11.3–14.5 points on every event. To summarize, narrative-network structure ranks next-day emergent relations above chance, though this advantage is concentrated among the most frequently mentioned, highly connected entities.
\end{abstract}




\maketitle

\section{Introduction}

Narrative is one of humanity's oldest inventions, as fundamental to how we organize experience as fire or the wheel. The stories we tell shape how we interpret events, form worldviews, and act in daily life, and at scale they drive the dynamics of public discourse. This is especially visible in the news, where stories are not static. A storyline unfolds over days and weeks as new actors enter, alliances shift, and events trigger consequences. Anyone following such a story forms expectations about what happens next. Automating this expectation estimation, if accurate, could potentially be widely applicable during crisis responses such as protests, disasters, political upheaval and warfare broadly. Narrative networks, graphs whose nodes are the actors, actions, and entities of a story, make its structure explicit. This raises a natural question: does that structure carry information about future states of that story.

Computational Narratology has largely been retrospective, treating narrative as a static snapshot and analyzing events only after they have unfolded \cite{gerard2025telegram, keim2026conspiracy, lugos2026tracking, zhao2024narrative}. Such methods describe what has already happened well, but lack the machinery to anticipate where a story is going. Corman et al.~\cite{corman_gnn}, conjectured that narrative graphs could be generative and the interactions already present among a story's elements can predict relationships not yet observed. However, they demonstrate this on a single fictional story, and whether the narrative network structure anticipates new interactions in real, evolving narratives remains untested. Accordingly, we propose the research question to test whether narrative networks can predict evolving narratives: 
Can the structure of a narrative network predict how a narrative evolves, in particular the link structure between nodes (e.g., actors and locations) in the network? We cast this as link prediction on evolving web-derived networks of worldwide online news, a graph-modeling problem on the Web (Graph Algorithms and Modeling for the Web) that supports applications such as media monitoring and misinformation response.

    To answer these questions, we build narrative networks from news text using a large language model as a Semantic Edge Parser (SEP). Unlike conventional parsers that extract SVO triples sentence by sentence, the SEP reads each full article in a single pass, resolving coreference and normalizing entities, and transforms it into a set of actor-action-entity edges.
    We treat the narrative as a longitudinal object: each day we build a graph from a rolling window of recent coverage and predict the directed links that emerge on the following day. Emergent links are edges present at time $t+1$ but absent at time $t$, connecting nodes that exist at both time steps. The node set is unchanged; only the relational structure evolves. We further emphasize that forecasting the arrival of entirely new elements is a distinct problem, and we do not address it here because a structural predictor can only score a pair once both endpoints are already in the graph; a brand-new element has no edges to score from. 
    
    We score candidate links with standard structural predictors, including Common Neighbors\cite{PhysRevE.64.025102}, Adamic-Adar\cite{adamic2003friends}, and Preferential Attachment~\mbox{\cite{libennowell2007link}}. We compare against EdgeBank\cite{poursafaei2022betterevaluationdynamiclink}, a memorization baseline, and a Temporal Graph Network (TGN) ~\mbox{\cite{rossi2020temporalgraphnetworksdeep}}, a graph neural network, that learns node representations from the time-ordered sequence of dated edges and serves as a strong state-of-the-art baseline for dynamic link prediction. Throughout, we separate emergent links from recurring ones. Prior work shows that standard temporal evaluation is dominated by edges that simply recur~\mbox{\cite{poursafaei2022betterevaluationdynamiclink}}, which can hide whether a method captures genuine narrative development.

    We test and evaluate our pipeline on three real events that vary in both domain and time period: the 2022 Freedom Convoy, the 2025--26 Iranian protests, and Hurricane Melissa (2025). The hurricane dataset differs in domain from the two protests, which guards against findings that hold for only one type of narrative. In this paper, we make three contributions. First, we extend Corman et al.'s narrative-graph formalism from a single static story to real, longitudinal narratives built from news with an LLM-based Semantic Edge Parser. Second, we formulate narrative evolution as an emergent link-prediction task, with a forward-chaining protocol that explicitly separates emergent links from recurring ones. Third, we study the emergent setting, in which we predict genuinely new links against randomly sampled negatives. Across three events and three LLM extractors, structural predictors outperform a memorization baseline, and we analyze how far this advantage is explained by endpoint degree and hub structure.

\section{Related Work}


The representation of narratives, in the computational sense~\cite{piper2021narrative}, can be broadly classified by the scale of the analysis and the structure they impose on a story. Some smaller-scale analyses rely on modeling narratives as character-interaction networks~\cite{labatut2019character} or event-evolution graphs~\cite{li2018narrative}. Our research primarily focuses on analyzing narratives across the corpus, thereby confining our scope to methods that are more relevant to this scale: similarity-based representations (distributional embeddings and similarity graphs) and relation-based representations. 
Earlier work on deriving
structure from distributional similarity builds vector representations of words,
sentences, and documents and organizes them into timelines, tracks their drift over time~\cite{hamilton2016diachronic}, or clusters them into similarity graphs, as surveyed in~\cite{norambuena2023survey}; more recent work in this direction
uses LLMs to extract salient entities and relate their co-occurrence to
financial-market dislocations~\cite{miori2025narratives}. Such representations
capture how entities relate in a corpus, for instance, which actors are similar.
However, their edges record only similarity or co-occurrence, not that one actor
acted on another. A related corpus-scale tradition uses topic models and media-framing analysis to characterize discourse~\cite{dimaggio2013exploiting, card2015media}, though these describe documents rather than relations among narrative elements.

In parallel, another line of work grounded in computational linguistics derives
entities (nodes) and relations (edges) from the sentence itself, extracting
subject--verb--object or semantic-role structure so that actors and entities
become the nodes and the verb becomes the relation between them. This idea runs
through quantitative narrative analysis, open information extraction, semantic
role labeling, and Abstract Meaning Representation, predating language models and
continuing into them~\cite{franzosi2010qna, carley1993coding,  sudhahar2015network, delcorro2013clausie,
ash2023relatio, pournaki2025narrative, tangherlini2020automated},
and it underpins much work on building narrative networks. While the study of
narrative networks spans NLP, sociology, political science, and the digital
humanities~\cite{bearman2000nazi, sudhahar2015network, labatut2019character},
most of this work represents narrative elements as nodes and actions as edges.
Corman et al.'s formalism~\cite{corman_gnn} takes a different approach, promoting
verbs to nodes so that structural measures such as degree and centrality can
gauge the importance of entities as well as actions. Promoting relations to nodes in this way connects to hypergraph and $n$-ary knowledge-graph representations~\cite{fatemi2020knowledge, rosso2020beyond}, which likewise represent relations as entities in their own right; we forecast over such a representation rather than only modeling it. We build on this
representation because it makes narrative structure explicit and lets us frame
narrative evolution as link prediction, a well-studied task with established
structural predictors.

A significant body of work predicts and analyzes how narratives evolve. Prior
work forecasts the next event over learned event chains, through the narrative
cloze~\cite{chambers2008chains, chambers-jurafsky-2009-unsupervised}, later extended with neural event models~\cite{granrothwilding2016whathappens, pichotta2016scripts}, and event-evolution graphs~\cite{li2018narrative}.
Both operate over event chains abstracted away from the particular story in which
they occurred, learning which event typically follows which across a genre, rather
than forecasting the emergent relations within a single unfolding narrative. A
separate line of research studies narrative change in a specific unfolding story,
but does so retrospectively through change-point detection over per-period
narrative networks~\cite{zhao2024narrative}, embedding
drift~\cite{gerard2025telegram, keim2026conspiracy}, semantic drift and
dispersion~\cite{lugos2026tracking}, and, where structure is examined, properties
of the participants rather than the story~\cite{nefriana2026leader,
gerard2026bridging}. Each characterizes a shift only after the corpus is complete. Similarly, work on tracking stories through news over time organizes coverage into timelines and story chains~\cite{allan1998tdt, shahaf2010connecting, shahaf2012trains}, but describes rather than forecasts their development.
The field's own survey places narrative forecasting outside its scope, noting that
predicting the next events ``lies between extraction and
generation''~\cite{norambuena2023survey}. We ask whether the structure present in
a narrative at one moment anticipates the relations that appear at the next.

We pose narrative evolution as predicting new edges between narrative elements, a
task that sits at the intersection of network science and computational
narratology. Classical link prediction ranks candidate pairs by structural
proximity, scoring each unobserved edge by how closely its endpoints sit in the
graph~\cite{libennowell2007link, lu2011link, adamic2003friends, zhou2009predicting,
tong2006fast}. Corman et al.'s analysis of the Three Little Pigs uses Random Walk
with Restart, a classical scorer, to surface edges that could plausibly appear
next~\cite{corman_gnn}. We adopt these classical scorers unchanged and adapt them
to the directed narrative graph, and introduce no new predictor. A central problem in link prediction, and in forecasting evolving narratives, is that models tend to predict recurring edges over the genuinely new ones that advance the story. This problem surfaces
particularly on the interaction streams learned by temporal graph networks such as
JODIE, DyREP, TGAT, and TGN~\cite{kumar2019jodie, trivedi2019dyrep, xu2020tgat, rossi2020temporalgraphnetworksdeep}, whose test
sets are dominated by recurring edges~\cite{poursafaei2022betterevaluationdynamiclink}.
Junuthula et al. therefore argue that dynamic link prediction should separate the
prediction of new links from that of previously observed ones, since a single
aggregate score conflates the two~\cite{junuthula2016evaluating}. A parallel line forecasts future facts in temporal knowledge graphs built from event streams~\cite{trivedi2017knowevolve, jin2020renet}, but over a fixed schema of entity and relation types rather than the open narrative elements we extract.

We build on the established work distinguishing new and recurring edges to design
our evaluation protocols and baselines~\cite{poursafaei2022betterevaluationdynamiclink,
junuthula2016evaluating}. Several recent works predict social
ties, narrative adoption, or cross-platform propagation rather than relations among
the narrative elements themselves~\cite{gu2026frame, gerard2025crossplatform,
amure2025templates}. Corman et al.'s narrative-network formulation already forecasts
new relations~\cite{corman_gnn}, so we additionally analyze its forecasting ability
on new versus recurring edges. Since we use LLMs for SEP, we treat the SEP parser as
a measurement instrument and report every result under three independent models, so
that no finding hinges on a single extractor~\cite{pangakis2025humans}.

\section{Methods}

We define a narrative network as a directed graph whose nodes are the narrative elements of a news story, its actors, actions, and entities, and whose edges record the direction of interaction between them. Our formulation differs from that of Corman et al.~\cite{corman_gnn} in that we do not assign type labels to these elements, whereas their method classifies nodes into Locations, Events, Actions, Participants, and Things. This section formalizes that graph, describes the Semantic Edge
Parser that builds it from text, defines the emergent link-prediction
task, and specifies the directed indices and evaluation protocol we
use to score it.
Our aim is to test whether the structure of a narrative network forecasts emergent interactions between narrative elements, not to propose new predictors. 
Although the structural scorers are defined for undirected graphs, we adapt them to the directed setting given the directed nature of our narrative networks. Table~\ref{tab:indices} gives the directed form of each scorer.

\subsection{Semantic Edge Parsing}
\label{sec:sep}



The first step in constructing the narrative networks is to extract the narrative graph from news articles with Semantic Edge Parsing (SEP), which decomposes each
sentence into directed, labeled edges of the form
(\emph{source}, \emph{target}, \emph{label}, \emph{weight}). The label gives the
syntactic role of the relation (subject, object, adjective) or, for prepositional relations, the preposition itself. For
example, ``The police cleared protesters from the bridge'' yields
\texttt{police}$\rightarrow$\texttt{clear} (subject),
\texttt{clear}$\rightarrow$\texttt{protester} (object), and
\texttt{clear}$\rightarrow$\texttt{bridge} (preposition \emph{from}), so that the
action \emph{clear} becomes a node linking the entities it relates. Promoting
actions to nodes rather than treating them as edge labels is the design choice
that lets structural measures apply to what happens in a story and not only to
whom it happens to.

\begin{figure}[tp]\centering
  \includegraphics[width=\columnwidth]{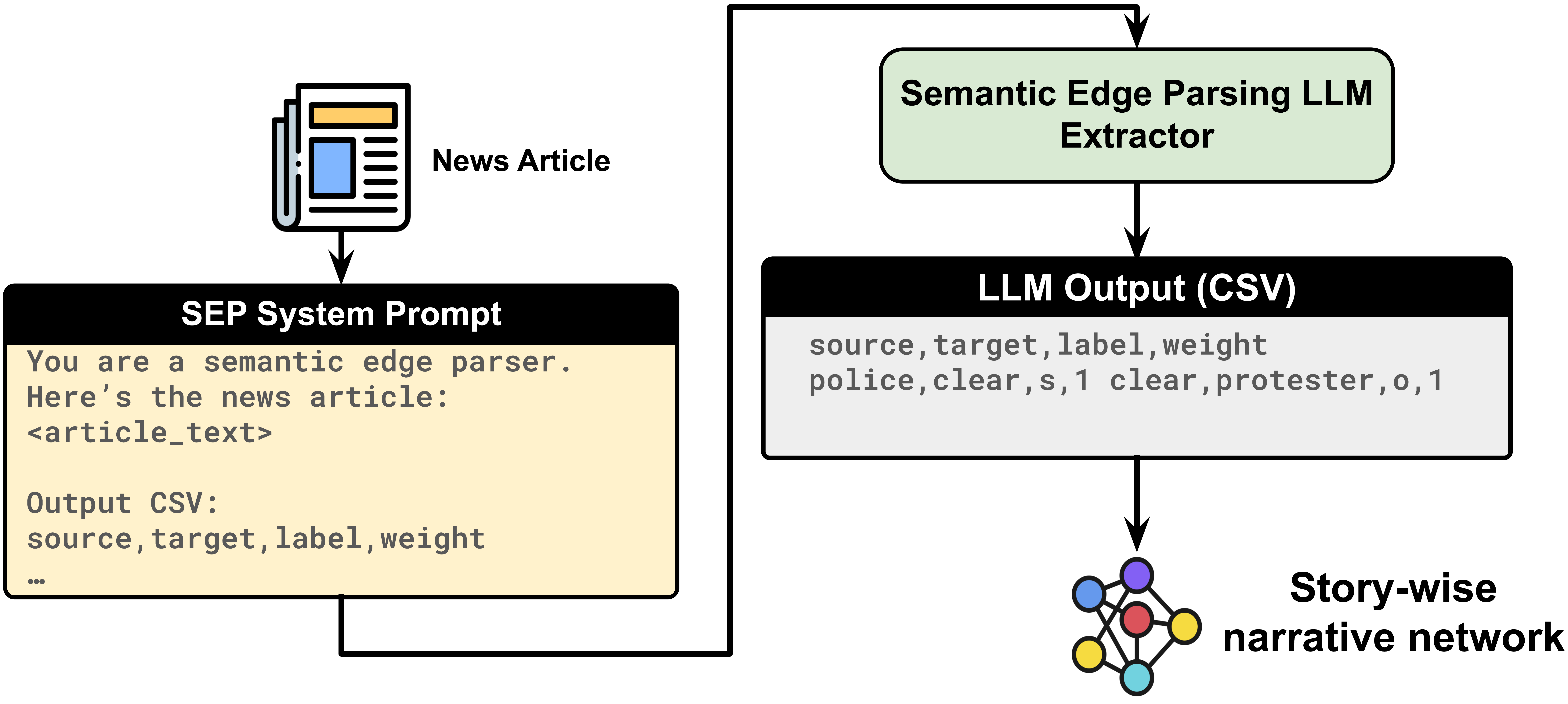}
   \caption{Semantic Edge Parsing on a single article. The SEP extractor takes
  the article text and, in a single LLM pass, emits a CSV edge list
  (\texttt{source, target, label, weight}), promoting each action to a node,
  resolving coreference (pronouns and short names map to their full entity), and
  normalizing tokens (lowercasing, infinitives, singular nouns). The returned
  edges form the article's story-wise narrative graph.}
  \label{fig:sep}
\end{figure}


We perform SEP with a large language model, which in a single pass resolves
coreference by replacing pronouns with the entities they denote and treats
multi-word expressions such as ``justin trudeau'' and ``canadian government'' as
single nodes. We parse each article independently, yielding a story-wise narrative graph, and pool these edge lists into the day and window level graphs mentioned in Section~\ref{sec:construction}. We note that the
parser is an instrumentation rather than a contribution and instead of validating it against a single ground truth, we report all results under three independent extractors (Section~\ref{sec:robustness}) and show that the findings do not depend on the choice of model.

\subsection{The Narrative Network}
\label{sec:construction}


We represent the narrative of a news story as a directed graph
$g = (V, E)$, where $V$ is a vertex set of narrative elements and $E$ is a set
of directed edges between them.

\begin{figure}[t]\centering
  \includegraphics[width=\columnwidth]{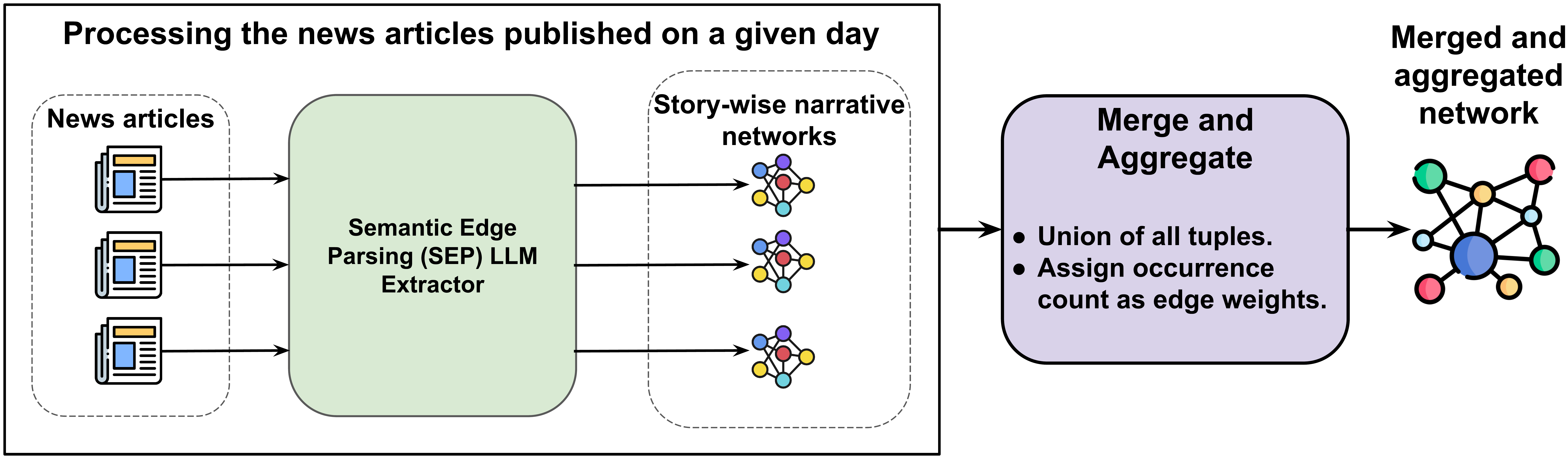}
   \caption{Merge and aggregation. The story-wise graphs of all articles published
  on a given day are unioned and identical edges are weighted by their occurrence
  count, producing the day-level network $g_t$. The same operation aggregates the
  daily snapshots $g_i$ into the windowed graph $G_{[t-W+1,\,t]}$.}
  \label{fig:merge}
\end{figure}

Let $a_{i,t}$ denote the $i$-th article published on day $t$, and let $N_t$ be the number of articles published on that day. A semantic edge parser (see Section~\ref{sec:sep})
maps each article to a set of weighted edges $E_{i,t}$, where each edge $(u, v, w)$ records a relation of weight $w$ between entities $u$ and $v$ extracted from the article. The edge set for day $t$ is the union of the per-article edge sets, $$E_t = \bigcup_{i=1}^{N_t} E_{i,t},$$ where edges with the same endpoints are merged. For every edge, we define the edge weight as count of duplicate edges with $u$ as the source node and $v$ as the destination node. The daily snapshot is the graph $g_t = (V_t, E_t)$, with $V_t$ the set of entities appearing as an endpoint of some edge in $E_t$. To capture a narrative that unfolds over several days, we aggregate daily snapshots
over a window of $W$ consecutive days ending at day $t$,
$$G_{[t-W+1,\,t]} = (V_{[t-W+1,\,t]}, E_{[t-W+1,\,t]}) = \bigcup_{i=t-W+1}^{t} g_i,$$
where the union of graphs takes the union of their vertex sets and merges
their edge sets as above, where the new weights are again the number of duplicates across all articles and days.
Both steps apply the same merge operation (Figure~\ref{fig:merge}): first
pooling articles into a day, then pooling days into a window.

For completeness, we include information on how we build the edge weights for the graph while describing the graph construction process. The structural scorers (Table~\ref{tab:indices}) we employ in Section~\ref{sec:task} do not require the edge weights and we therefore do not use edge weights for our analysis. However, we do provide an ablation study in Appendix~\ref{app:weighted} to investigate whether edge weight would make any difference in forecasting new links and observe that weighting the graph does not improve the forecasting performance. 


\subsection{Emergent Link Prediction}
\label{sec:task}



We define emergent link prediction on the window graph $G_{[t-W+1,\,t]}$, built
from the news articles in the $W$-day window ending at day $t$, as predicting the
directed edges that appear at $t+1$ whose endpoints already lie in
$G_{[t-W+1,\,t]}$ but which are themselves absent from it. We denote this set of
emergent edges by $\Delta_{t+1}$ and define it as:




\begin{equation}
\label{eq:emergent}
\begin{split}
\Delta_{t+1} = \bigl\{\, (u,v) \in E_{t+1} : {} &(u,v) \notin E_{[t-W+1,\,t]}, \\
&u,v \in V_{[t-W+1,\,t]} \,\bigr\}.
\end{split}
\end{equation}

Restricting to nodes already present
separates the emergence of new \emph{relations} from the arrival of new
\emph{elements}, which is a different phenomenon. We also note that our emphasis is on the new relations in $\Delta_{t+1}$, not on the links that recur at $t+1$. We target links that are absent from the $W$-day window but appear at $t+1$, and exclude links already present in the window. Because such recurring links are never candidates in the emergent task, they neither help nor hurt the structural scorers defined in Table~\ref{tab:indices}. We instead evaluate them under the standard recurrence-inclusive protocol.




\textbf{Structural Scorers.} 
Link prediction ranks the graph’s missing candidate pairs by structural proximity, using properties such as endpoint degree and shared-neighbor overlap.
Conventionally, the standard structural scorers operate only on undirected graphs.
Another issue with the scorers is that they are symmetric and assign the same score
to the links $(u,v)$ and $(v,u)$. This conflicts with our narrative-network
construction, because the network contains directed edges, and symmetrizing $(u,v)$
and $(v,u)$ would lose the information on who did what to whom, which is integral to
analyzing narratives. We therefore redefine the structural scorers (Table~\ref{tab:indices})
to operate on the directed narrative network. Each scorer is built from the neighborhoods of
the pair $(u, v)$, so we first recast those for the directed setting. Let $N^{+}(u)$ and $N^{-}(u)$ denote the out- and in-neighborhoods of a node $u$
in $G_{[t-W+1,\,t]}$, with out-degree $d^{+}(u) = |N^{+}(u)|$, in-degree
$d^{-}(u) = |N^{-}(u)|$, and total degree $d(u) = d^{+}(u) + d^{-}(u)$.
The common neighbors of an ordered pair $(u,v)$ are the nodes on a directed
two-path $u \to n \to v$ and can be defined as:

\begin{equation}
  \Gamma(u,v) = N^{+}(u) \cap N^{-}(v).
\end{equation}



so that $n$ relays $u$ to $v$. This is asymmetric because
$\Gamma(u,v) \neq \Gamma(v,u)$ in general. Table~\ref{tab:indices} provides the six
scorers under this definition. Random Walk with Restart is natively directed and therefore requires no adaptation.
Each scorer assigns every candidate pair a score, with a higher score indicating a greater likelihood of a next-day edge, and ranks the candidates accordingly.

%
\begin{table}[t]\centering\small
\caption{Structural indices adapted to the directed setting.
$\Gamma(u,v)$ is the set of nodes on a directed two-path $u \to n \to v$; $d^{+}$ and $d^{-}$ are out- and in-degree; $d = d^{+} + d^{-}$.}
\begin{tabular}{ll}
\toprule
Index & Directed form \\
\midrule
Common Neighbors        & $|\Gamma(u,v)|$ \\
Adamic--Adar            & $\sum_{n\in \Gamma(u,v)} 1/\log d(n)$ \\
Resource Allocation     & $\sum_{n \in \Gamma(u,v)} 1/d(n)$ \\
Jaccard                 & $|\Gamma(u,v)| \,/\, |N^{+}(u) \cup N^{-}(v)|$ \\
Hub-Depressed           & $|\Gamma(u,v)| \,/\, \max\{d^{+}(u),\, d^{-}(v)\}$ \\
Preferential Attachment & $d^{+}(u)\, d^{-}(v)$ \\
\bottomrule
\end{tabular}

\label{tab:indices}

\end{table}

\subsection{Data}
\label{sec:data}
 
We investigate the validity of narrative networks on three events drawn from
different domains and time periods: the 2022 Freedom Convoy protests, the 2025
Hurricane Melissa, and the 2025--26 Iranian protests. The Freedom Convoy and the
Iranian protests belong to the same domain of civil protest, while Hurricane
Melissa is a natural disaster, so the three together let us check that our
findings are not specific to a single kind of story. 
 
All three corpora were collected from the GDELT Project~\cite{leetaru2013gdelt}
through the same pipeline. For each event we queried GDELT's Global Knowledge
Graph over the event's date window with an event-specific filter: the Freedom
Convoy by \emph{convoy} and \emph{trucker} keywords restricted to articles
located in Canada, Hurricane Melissa by the storm's name, and the Iranian
protests by \emph{protest}, \emph{unrest}, and \emph{crackdown} themes localized
to Iran. We retrieved the full body text of each returned URL with trafilatura \cite{barbaresi-2021-trafilatura} 
and removed near-duplicate articles by deduplicating on the article text, which collapses syndicated wire copy. We report the summary statistics for articles and resulting graphs in Table~\ref{tab:corpus}. 
 The graph statistics in this table represent the values from the Haiku extractor; the other extractors
are comparable.
 
\begin{table}[t]\centering\small
\setlength{\tabcolsep}{2pt}
\caption{Corpus statistics for the three events (Haiku extractor). ``Articles''
counts articles contributing at least one relation. ``Total edge instances'' is
the sum of edge weights, i.e.,\ total relation mentions. Surprise is the fraction
of test edges never seen in training, and reoccurrence the fraction seen in
training, on a chronological 85/15 split over unique directed edges (the two sum
to one); we report them so our datasets can be placed alongside the dynamic
link-prediction benchmarks of Poursafaei et
al.~\cite{poursafaei2022betterevaluationdynamiclink}.}\label{tab:corpus}
\begin{tabular}{lrrr}\toprule
 & Convoy & Iran & Melissa\\ \midrule
Period & 2022 & 2025--26 & 2025\\
Articles & 5,417 & 7,763 & 5,366\\
Days & 70 & 59 & 45\\
Articles/day (mean) & 77 & 132 & 119\\
Articles/day (range) & 2--368 & 10--599 & 7--968\\
Unique nodes & 64,068 & 110,302 & 57,865\\
Unique directed edges & 463,247 & 866,419 & 361,534\\
Total edge instances & 1,200,189 & 2,273,081 & 5,589,020\\
Forward-chaining eval days & 38 & 37 & 23\\
Surprise (unseen test edges) & 0.647 & 0.659 & 0.726\\
Reoccurrence (train edges in test) & 0.353 & 0.341 & 0.274\\
\bottomrule
\end{tabular}\end{table}
 
 
\subsection{Evaluation Protocol}
\label{sec:protocol}

\begin{figure*}[t]\centering
  \includegraphics[trim=0 25 0 25, clip,  width=0.9\textwidth]{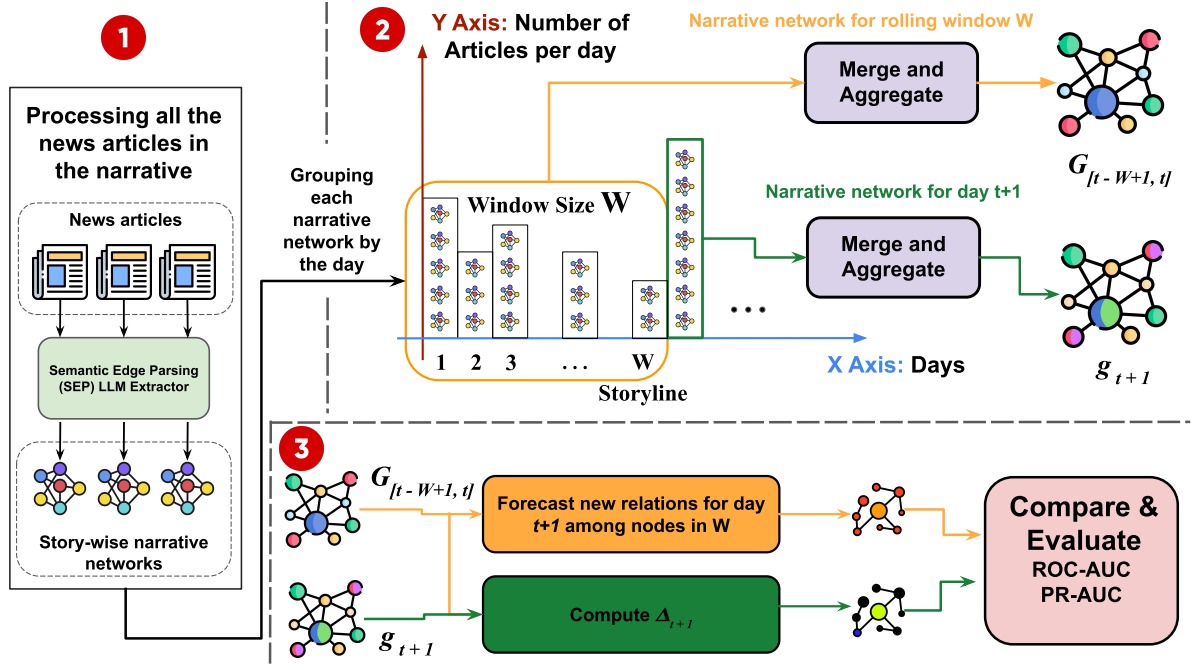}
  \caption{Evaluation pipeline. Per-article graphs are merged into daily
snapshots $g_t$, which are merged into the windowed graph $G_{[t-W+1,\,t]}$.
The emergent edge set is $\Delta_{t+1}$ defined in Eq~\ref{eq:emergent}.
Each method scores candidate pairs in the window, and the scores are evaluated
against $\Delta_{t+1}$ by ROC-AUC and PR-AUC.}
  \label{fig:pipeline}
\end{figure*}

Figure~\ref{fig:pipeline} summarizes the evaluation pipeline. We evaluate
under two protocols: emergent and standard protocol. The \emph{emergent} protocol is our primary setting. It
is training-free, uses forward chaining over the horizon, and evaluates only
relations that have not previously appeared. The \emph{standard} protocol is
a chronological split in which test edges may recur from training. We include it for two reasons: it is the conventional recurrence-inclusive
evaluation, and it accommodates TGN\cite{rossi2020temporalgraphnetworksdeep}, the one baseline that must be trained. The
protocols differ fundamentally in evaluation objective, negative ratio, and
chronological split, so scores are comparable only within a protocol. Across the
protocols we compare only the ordering of methods and not their scores.

\textbf{Emergent protocol.} For each day $t$ we construct the windowed graph
$G_{[t-W+1,\,t]}$ from the preceding $W$ days and define the emergent edge set as specified in Eq.~\ref{eq:emergent}.
Candidate pairs are the
unconnected ordered pairs among the 1,000 highest-degree nodes in the
window, after removing a 25-word stop-list of light verbs and generic nouns (Appendix~\ref{app:hyper}). A candidate is
positive if it lies in $\Delta_{t+1}$. We retain all positives and sample five
negatives per positive uniformly from the remaining candidates (random negatives), giving a
positive rate of approximately $0.167$; this follows the random negative
sampling of \citet{poursafaei2022betterevaluationdynamiclink}. Each method
assigns a score to every candidate, and we evaluate the scores against the
labels. We then advance one day and repeat. We set $W=14$ and report a sweep
over $W$ in Appendix~\ref{app:window}. We refrain from including days whose window has fewer than 50 nodes or fewer than 10 positive edges, since per-day ROC-AUC and PR-AUC are unstable when positives are scarce. 
Table~\ref{tab:corpus} reports the number of evaluation days per event.
The 1,000-node restriction keeps the
candidate space tractable and matches the node set used in the standard
protocol. Section~\ref{sec:core} reports an ablation over this cut-off.

\textbf{Standard protocol.} We split the narrative trajectory chronologically
into training (70\%), validation (15\%) and test (15\%) sets with $1:1$
negatives, following standard dynamic link-prediction
evaluation~\cite{poursafaei2022betterevaluationdynamiclink}. We run all methods
under this protocol with DyGLib~\cite{yu2023towards} on the same
top-1,000-node graph.

\textbf{Baselines.} Preferential Attachment scores a pair by the product of its
endpoint degrees in the window. EdgeBank~\cite{poursafaei2022betterevaluationdynamiclink}
is a training-free memorization baseline that predicts an edge if it has been
seen before; we use its unlimited-memory configuration, which is its strongest.
Random Walk with Restart is the method of \citet{corman_gnn}. TGN is a temporal
graph neural network that maintains node memory over the interaction
stream~\cite{rossi2020temporalgraphnetworksdeep}. Preferential Attachment,
EdgeBank, Random Walk with Restart and a chance baseline (ROC-AUC $=0.5$) run
under both protocols. TGN requires a single large training split, so it runs
under the standard protocol only, and we make no claim about TGN on the
emergent protocol.

\textbf{Metrics.} We report ROC-AUC and PR-AUC (average precision) averaged
over evaluation days. We test the
difference between Preferential Attachment and EdgeBank with a Wilcoxon
signed-rank test on the paired per-day scores.

The evaluation pipeline (Figure~\ref{fig:pipeline}) consists of three stages. First, we merge the daily snapshots $g_{t-W+1}, \ldots, g_t$ into the windowed graph $G_{[t-W+1,\,t]}$. Second, we apply a link prediction method to this graph to score and rank the candidate edges. Third, we compare the ranked candidates against the emergent link set $\Delta_{t+1}$ using ROC-AUC and PR-AUC metrics.

\section{Results}

\subsection{Emergent Link Prediction}



\begin{table}[t]\centering\footnotesize
\setlength{\tabcolsep}{3pt}
\caption{Emergent link prediction: ROC-AUC / PR-AUC, mean over three LLM
extractors (Haiku, GPT-4.1, Gemini), rolling window $W=14$, $5:1$ negatives
(chance ROC-AUC $=0.5$, positive base rate $\approx0.17$). Preferential
Attachment is the strongest predictor and beats the EdgeBank memorization
baseline on every event (Wilcoxon $p<10^{-6}$; PA wins on every day-split).}
\begin{tabular*}{\columnwidth}{@{\extracolsep{\fill}}lcccccc@{}}
\toprule
 & \multicolumn{2}{c}{Convoy} & \multicolumn{2}{c}{Iran} & \multicolumn{2}{c}{Melissa}\\
\cmidrule(lr){2-3}\cmidrule(lr){4-5}\cmidrule(lr){6-7}
Scorer & ROC & PR & ROC & PR & ROC & PR\\
\midrule
Common Neigh.     & 0.737 & 0.395 & 0.721 & 0.362 & 0.689 & 0.358\\
Adamic--Adar      & 0.740 & 0.414 & 0.723 & 0.377 & 0.690 & 0.380\\
Resource Alloc.   & 0.737 & 0.404 & 0.719 & 0.368 & 0.686 & 0.358\\
Jaccard           & 0.687 & 0.294 & 0.674 & 0.285 & 0.659 & 0.253\\
Hub-Depressed     & 0.675 & 0.278 & 0.661 & 0.269 & 0.657 & 0.250\\
Pref.\ Attachment & \textbf{0.785} & \textbf{0.456} & \textbf{0.755} & \textbf{0.405} & \textbf{0.793} & \textbf{0.485}\\
RWR (Corman)      & 0.703 & 0.329 & 0.681 & 0.303 & 0.727 & 0.367\\
EdgeBank          & 0.640 & 0.313 & 0.642 & 0.298 & 0.679 & 0.367\\
\bottomrule
\end{tabular*}

\label{tab:emergent}
\end{table}

\begin{figure}[htbp]\centering
  \includegraphics[width=\columnwidth]{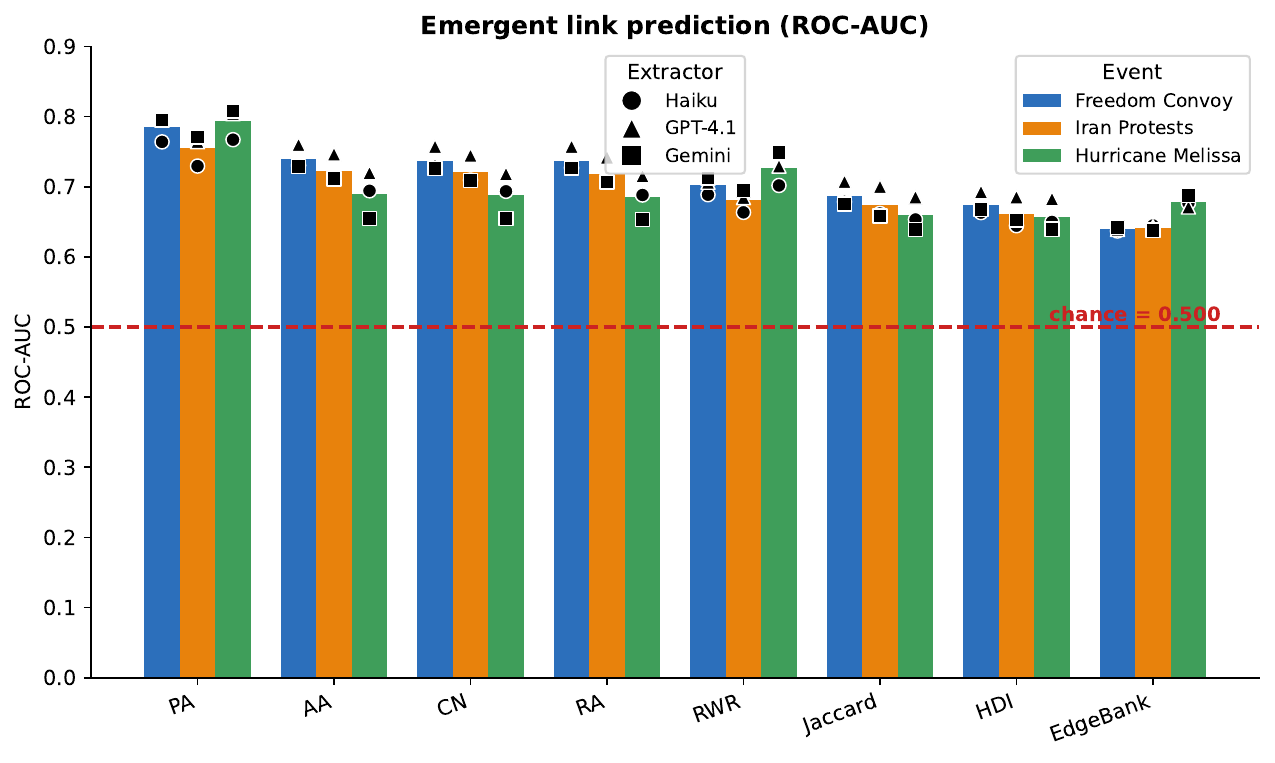}
  \caption{Emergent link prediction, ROC-AUC. Bars are the mean over the three SEP
  extractors; markers show per-extractor values (circle Haiku, triangle GPT-4.1,
  square Gemini). PA is strongest on every event and beats EdgeBank.}
  \label{fig:results_roc}
\end{figure}

\begin{figure}[t]\centering
  \includegraphics[width=\columnwidth]{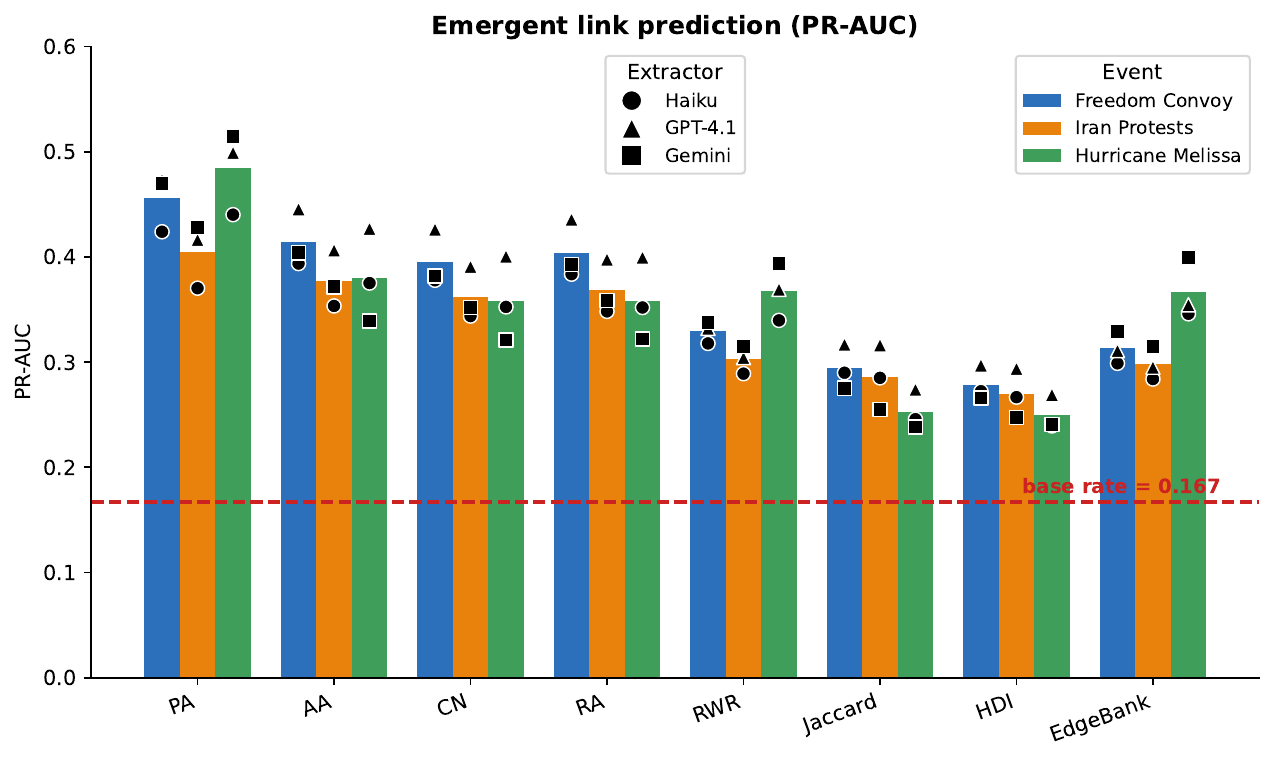}
  \caption{Emergent link prediction, PR-AUC (positive base rate $0.167$). Same
  bars and markers as Fig.~\ref{fig:results_roc}.}
  \label{fig:results_pr}
\end{figure}

Table~\ref{tab:emergent} reports ROC-AUC and PR-AUC for every scorer on the
three events, averaged over the three LLM extractors; Fig.~\ref{fig:results_roc} and~\ref{fig:results_pr} 
plot the same result. Preferential Attachment is the strongest scorer on all
three events, with a mean ROC-AUC of 0.785 on the Freedom Convoy, 0.755 on the
Iran protests, and 0.793 on Hurricane Melissa, and PR-AUC of 0.456, 0.405, and
0.485 against a positive base rate of 0.167. It exceeds the EdgeBank
memorization baseline by 14.5, 11.3, and 11.4 ROC-AUC percentage points on the three
events. The local neighborhood scorers rank next. On the
Convoy and Iran events, Adamic--Adar, Common Neighbors, and Resource Allocation lie between 0.719 and 0.740 ROC-AUC and differ from one another by at most 0.004, so we treat them as tied. The
normalized indices, Jaccard and Hub-Depressed index,
are the weakest scorers overall (0.657 to 0.687 ROC-AUC). This pattern suggests that 
the emergent signal is carried by raw degree and hub structure of the window
graph rather than by normalized overlap, which is consistent with Preferential
Attachment, a degree-only scorer, ranking first on every event.

Two scorers behave differently on Hurricane Melissa. Random Walk with Restart,
the method used by Corman et al.~\cite{corman_gnn}, trails the local scorers on
the Convoy (0.703) and Iran (0.681) but ranks second on Melissa
(0.727). EdgeBank is the weakest scorer on the
Convoy (0.640) and Iran (0.642) but exceeds Jaccard and Hub-Depressed on Melissa (0.679). Neither scorer approaches
Preferential Attachment on any event, so the ordering at the top of the
ranking is stable across events.

\subsection{Emergent Versus Standard Evaluation}

\begin{table}[t]\centering\small
\setlength{\tabcolsep}{3pt}
\caption{Link prediction on the \emph{standard} recurrence-inclusive protocol
across all three events (chronological 70/15/15 split, $1:1$ negatives; each
cell is ROC-AUC / PR-AUC). With recurring edges included, memorization (EdgeBank)
and the temporal graph network (TGN) dominate the structural heuristics on every
event. EdgeBank and TGN are run with DyGLib on the same
top-1,000-node graph.}
\begin{tabular}{lccc}\toprule
Method & Convoy & Iran & Melissa\\ \midrule
Adamic--Adar     & 0.660/0.659 & 0.639/0.639 & 0.672/0.667\\
Common Neigh.    & 0.658/0.657 & 0.638/0.638 & 0.669/0.663\\
Pref.\ Attach.   & 0.654/0.657 & 0.637/0.641 & 0.659/0.657\\ \midrule
EdgeBank         & \textbf{0.781}/0.716 & \textbf{0.758}/0.692 & 0.785/0.727\\
TGN              & 0.743/\textbf{0.752} & 0.736/\textbf{0.752} & \textbf{0.817}/\textbf{0.818}\\
\bottomrule
\end{tabular}

\label{tab:standard}
\end{table}

\begin{figure}[tb]\centering
\includegraphics[width=\columnwidth]{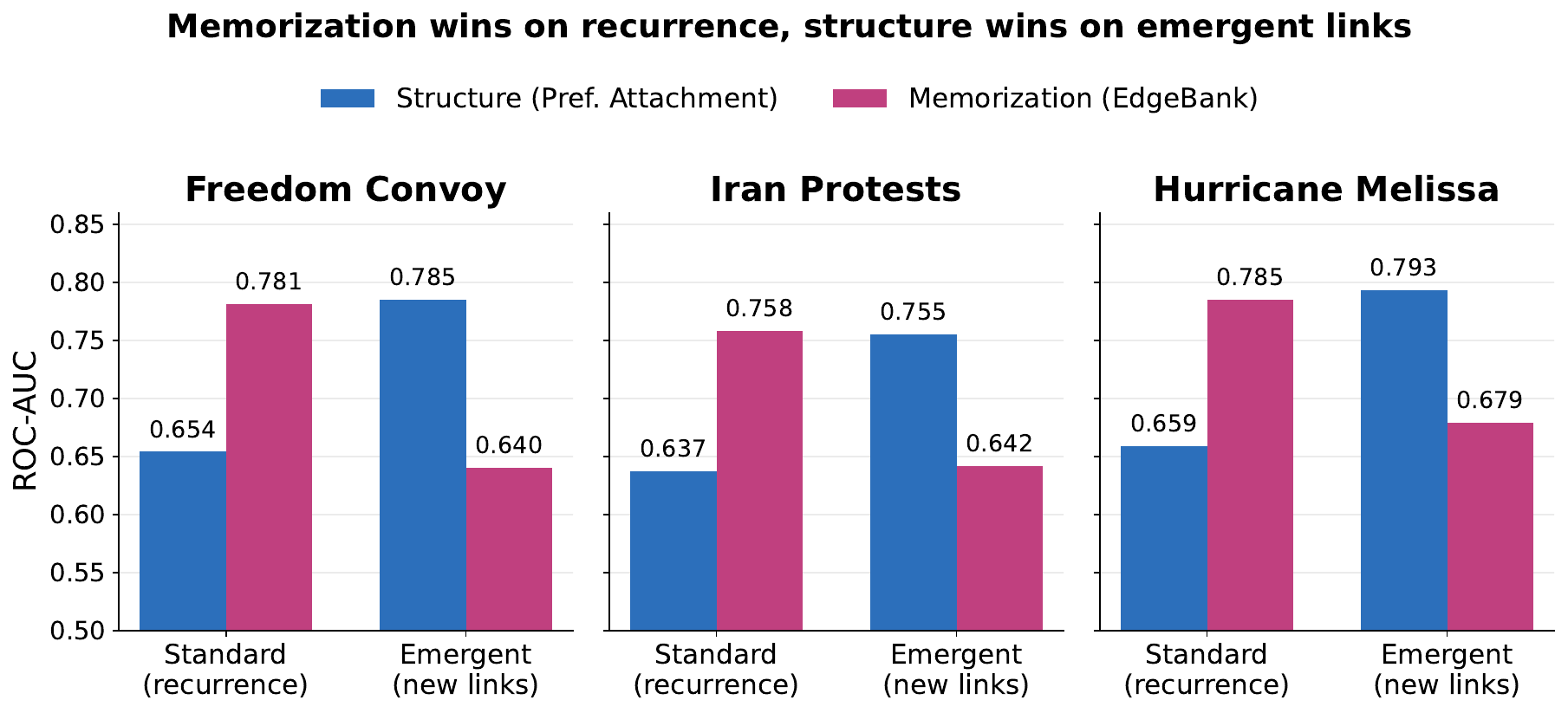}
\caption{Memorization matches structure on the recurrence-inclusive protocol
but collapses on emergent links, where structural prediction holds.}
\label{fig:emergent}
\end{figure}

The separation above is specific to emergent links. Table~\ref{tab:standard}
evaluates the recurrence-inclusive standard protocol on all three events, where
the test set includes edges that recur from training. Under this protocol
memorization and a temporal graph network dominate the structural heuristics
everywhere: EdgeBank leads on the Convoy and Iran (0.781 and 0.758 ROC-AUC) and
TGN on Melissa (0.817), while the structural indices sit near 0.64 to 0.67.
Fig.~\ref{fig:emergent} contrasts the two settings: on the standard protocol
memorization leads (EdgeBank 0.781 against 0.654 for Preferential Attachment on the Convoy), but the ordering reverses on the emergent protocol, where EdgeBank collapses to 0.640 while Preferential Attachment rises to 0.785. Standard temporal link
prediction is thus dominated by recurrence, and the predictive value of
narrative structure becomes visible only once recurrence is removed, which is the
regime our forward-chained protocol targets.

\subsection{Why Structure Wins: A Case Study}

\begin{figure}[t]\centering
  \includegraphics[width=\columnwidth]{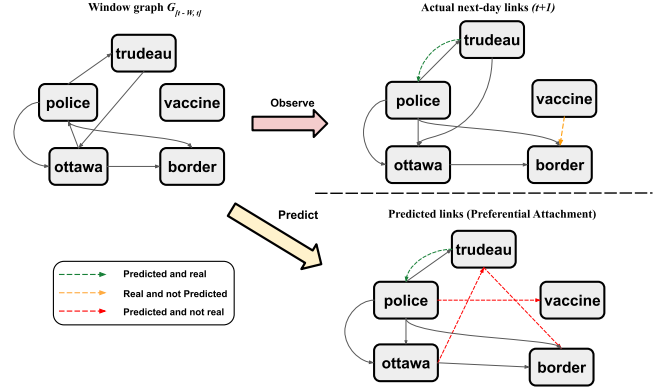}
  \caption{
  Why Preferential Attachment wins, on the Freedom Convoy for 2022-02-20 over a readable five-node subgraph of central actors. The window graph $G_{[t-W+1,t]}$ (left) is compared against the actual next-day relations (top) and PA's predictions (bottom). Grey edges are relations already present in the window; green is a correctly predicted new relation (\texttt{trudeau}$\to$\texttt{police}), orange a real new relation PA missed
  (\texttt{vaccine}$\to$\texttt{border}), and red are PA predictions that did not
  occur (\texttt{ottawa}$\to$\texttt{trudeau}, \texttt{trudeau}$\to$\texttt{border},
  \texttt{police}$\to$\texttt{vaccine}). PA's false positives all attach to the
  highest-degree hub, \texttt{trudeau}, consistent with its degree-driven
  mechanism.}
  \label{fig:casestudy}
\end{figure}

Figure~\ref{fig:casestudy} makes the mechanism concrete on the Freedom Convoy for
2022-02-20. Preferential Attachment correctly anticipates the emergent link
\texttt{trudeau}$\to$\texttt{police}, the crackdown relation the day's coverage
introduces, and its errors are equally telling. Every false positive it commits
(\texttt{ottawa}$\to$\texttt{trudeau}, \texttt{trudeau}$\to$\texttt{border},
\texttt{police}$\to$\texttt{vaccine}) attaches to \texttt{trudeau}, the
highest-degree node in the window, while the one real relation it misses
(\texttt{vaccine}$\to$\texttt{border}) sits between two low-degree, peripheral
nodes. This is precisely the behavior a degree-driven scorer should exhibit: it
concentrates its bets on hubs, where new relations most often form, at the cost of
over-predicting into those hubs and overlooking the periphery. The same logic
explains why memorization fails here. Across the three events, 59 to 68 percent of
emergent links are genuinely new, appearing for the first time anywhere on the day
they emerge (68\% on the Convoy, 63\% on Iran, and 59\% on Melissa). A memorization
baseline such as EdgeBank scores any never-before-seen pair at zero and is
therefore blind to the majority of emergent links, whereas endpoint degree stays
informative for them: a new relation between two active hubs is likely precisely
because both endpoints are active, regardless of whether that specific pair has
ever co-occurred. Structure wins on emergent links because most of them are new,
and novelty is where memorization has nothing to say and degree still does.

\begin{figure}[t]\centering
  \includegraphics[width=0.8\columnwidth]{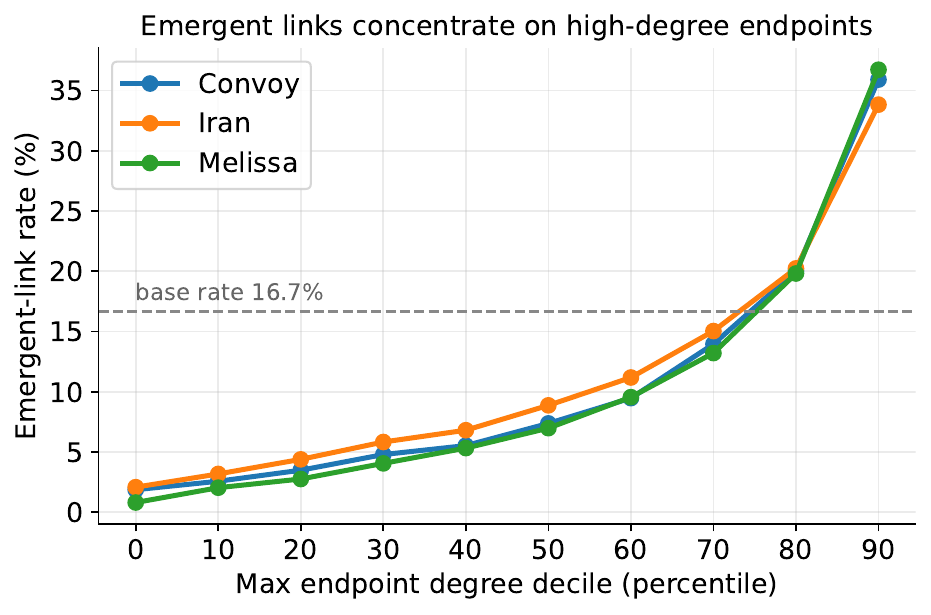}
  \caption{Emergent-link rate rises steeply with endpoint degree, from
  roughly 1--2\% in the lowest decile to 34--37\% in the top, on every event
  (base rate 0.167).}
  \label{fig:degree_stratified}
\end{figure}

\noindent\textbf{Emergence concentrates on hubs.} The probability that a
candidate pair becomes an emergent link rises steeply with endpoint degree,
from roughly 1--2\% in the lowest decile to 34--37\% in the top decile,
consistently across all three events (Fig.~\ref{fig:degree_stratified}).
Accordingly, 44--51\% of all emergent links have at least one endpoint in the
top degree decile, and the median endpoint-degree percentile of a true emergent
link is 0.88--0.90 versus 0.69--0.70 for sampled negatives.

\begin{figure}[t]\centering
  \includegraphics[width=0.8\columnwidth]{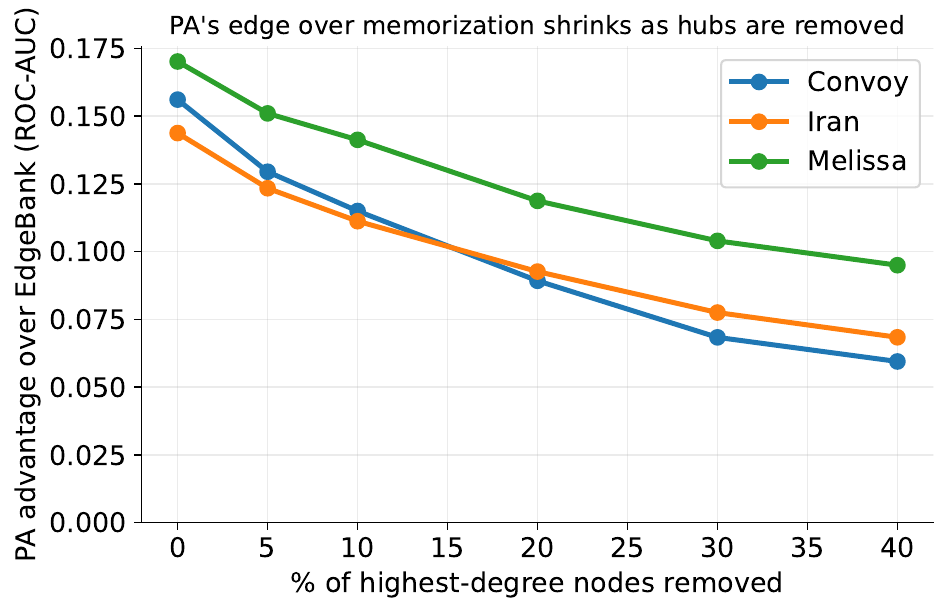}
  \caption{Hub-removal ablation: PA's margin over EdgeBank shrinks as
  the highest-degree nodes are progressively removed before scoring.}
  \label{fig:hub_removal}
\end{figure}

\noindent\textbf{Removing hubs removes the advantage.} To probe how much of PA's advantage is carried by the hubs, we progressively remove the highest-degree nodes
before scoring. PA's margin over EdgeBank shrinks monotonically as hubs are
stripped, from +0.156 with no removal to +0.059 once the top 40\% of hubs are
gone on the Convoy, and similarly on Iran (+0.143 to +0.068) and Melissa
(+0.170 to +0.095; Fig.~\ref{fig:hub_removal}).

\subsection{Robustness to the Candidate-Set Size}
\label{sec:core}
Table~\ref{tab:core} varies the candidate-set cutoff from 500 to 2,000 nodes
(Haiku extractor). Preferential Attachment is the top scorer at every cutoff on
all three events. Absolute scores rise with the cutoff, because a larger
candidate pool adds low-activity node pairs that endpoint degree separates
easily, but the ranking is invariant and PA's margin over the EdgeBank baseline
(which stays between 0.63 and 0.69) widens throughout. The main finding therefore
does not depend on the 1,000-node cutoff used elsewhere.
\begin{table}[t]\centering\small
\caption{Preferential Attachment ROC-AUC (Haiku extractor) as the candidate-set
cutoff varies. PA is the top scorer at every cutoff on every event; the EdgeBank
baseline stays in $[0.63,0.69]$ throughout.}
\setlength{\tabcolsep}{7pt}
\begin{tabular}{lcccc}\toprule
Event & 500 & 1,000 & 1,500 & 2,000\\ \midrule
Freedom Convoy    & 0.717 & 0.764 & 0.800 & 0.827\\
Iran Protests     & 0.686 & 0.731 & 0.764 & 0.790\\
Hurricane Melissa & 0.729 & 0.768 & 0.803 & 0.828\\
\bottomrule
\end{tabular}

\label{tab:core}
\end{table}

\subsection{Robustness Across LLM Extractors}\label{sec:robustness}
The markers in Figs.~\ref{fig:results_roc} and~\ref{fig:results_pr} show the values per-extractor behind each
bar. The ranking of scorers is stable across Haiku, GPT-4.1, and Gemini:
Preferential Attachment is the best scorer under every extractor on every event.
GPT-4.1 and Gemini generally give slightly higher scores than Haiku, but all
three sit well above EdgeBank on every event, so the ranking of scorers under random negatives does not depend on the choice of extractor.


\section{Conclusion and Discussion} 

In this study, we investigated whether the structure of a narrative network anticipates how a story continues, and posed it as forecasting the emergent links that later coverage introduces. In experiments across three events that vary in domain and time period, we find structural predictors rank next-day emergent relations above chance under random negatives. We further identified preferential attachment as the strongest and most consistent, exceeding the EdgeBank memorization baseline by 11 to 15 ROC-AUC percentage points on every event. A degree-based predictor leads, indicating that the signal is concentrated in high-degree, frequently mentioned entities rather than in finer relational structure.


We also observe that the forecasting signal is local rather than path-based, which is evident from the inferior performance of Random Walk with Restart compared to local scorers such as Common Neighbors, Adamic--Adar, and preferential attachment. This suggests that for emergent narrative relations the relevant structure is the activity of the endpoints rather than global proximity across the graph. While Corman et al.~\cite{corman_gnn} rely on Random Walk with Restart in their narrative-graph analysis, our results show that for emergent links in an evolving narrative, local endpoint structure (as captured by Preferential Attachment) is the stronger signal. We find that this advantage is specific to the emergent setting. Under the standard, recurrence-inclusive protocol, whose test set includes links that have already appeared earlier in the story along with the new ones, the ordering reverses and memorization and a temporal graph network take the lead, because this protocol is dominated by edges that simply recur. The memorization baseline keeps predicting the links it has already seen, so it excels when the test edges are recurring, as under the standard protocol, but has nothing to offer when they are genuinely new, as in the emergent task. This shows that the apparent advantage is specific to the emergent, random-negative setting.

\textbf{Limitations.} We predict relations between elements already present in the window and do not model the arrival of entirely new elements, which is a distinct cold-start problem. Our predictors are unsupervised structural indices rather than a model trained for the task, and our graphs are produced by a language model whose prompted co-reference is imperfect on hard cases such as long-range pronoun resolution. We mitigate extraction dependence by reporting every result under three independent extractors, but some effect of the extractor may still remain. We leave a human-validated evaluation of extraction accuracy to future work. We study three English-language events across two domains, and narratives that bifurcate into competing sub-stories may not be well described by a single accumulating graph. Each is a direction for future work, alongside learned temporal models for the emergent regime, representations that combine structure with content, and forecasting the arrival of new elements rather than only new relations among existing ones. Finally, our emergent-protocol results use uniformly random negatives, which favor degree-based scorers~\cite{poursafaei2022betterevaluationdynamiclink}; endpoint degree and news salience may account for much of Preferential Attachment's advantage, and isolating a genuinely structural effect would require historical or degree-matched negatives and learned temporal models on the emergent subset. These are directions for future work.

\section*{Ethics and Privacy Statement}
This work uses only public news indexed by GDELT and involves no human
subjects or private data. To mitigate bias from the LLM extractor, we
report every result under three independent extractors and treat the
parser as a measurement instrument rather than ground truth.

\bibliographystyle{ACM-Reference-Format}
\bibliography{acmart}

\appendix

\section{Semantic Edge Parsing Prompt}
\label{app:prompt}
The exact system prompt supplied to every extractor is reproduced below. It is
applied once per article; a single pass emits the CSV edge list and performs
coreference resolution and normalization.

\begin{sepprompt}
You are a semantic edge parser. When given text, apply Semantic Edge Parsing
(SEP) to produce an edgelist in CSV format with columns: source, target,
label, weight.

Edge Types
1. Subject edge (s): connects a noun to the verb it performs. noun -> verb.
2. Object edge (o): connects a verb to its direct or indirect object. verb -> noun.
3. Adjective edge (a): connects a modifying word to the noun it modifies. modifier -> noun.
4. Prepositional edge: the preposition (on, under, for, in, from, of, about,
   after, by, through, with, at, against) is the edge LABEL; verb -> noun or
   noun -> noun. The preposition is NEVER a source or target node.

Normalization Rules
* All words lowercase.
* Verbs reduced to infinitive form ("announced" -> "announce").
* Nouns reduced to singular form ("policies" -> "policy").
* Multi-word proper nouns kept as single nodes ("u.s. supreme court").
* Possessives become adjective edges ("Trump's policy" -> trump,policy,a).
* Be verbs included ("prices are high" -> prices,be,s and be,high,o).

Auxiliary Verbs
* Modals prepended with a hyphen ("must identify" -> "must-identify").
* Negation prepended with "not-" ("did not announce" -> "not-announce").
* Tense/aspect auxiliaries stripped; keep only the main verb infinitive
  ("was cleared" -> "clear", "has been shot" -> "shoot").
* Combine negation and modal ("could not pay" -> "cannot-pay").

Pronoun Coreference
* Replace ALL pronouns with the noun they refer to; no pronoun is ever a node
  (he, she, it, they, we, him, her, them, his, their, our, this, that, these,
  those). For quoted speech, resolve to who/what the speaker means; when
  ambiguous, use the most recently mentioned relevant noun.

Partial-name Coreference
* Resolve shortened references to the longest form in the text (both "Justin
  Trudeau" and "Trudeau" -> "justin trudeau"), only when context is unambiguous.

Other Rules
* Conjunctions produce separate edges for each element.
* Embedded clauses resolve the true subject and verb of each clause.
* Edge weight = number of times the same (source, target, label) triple appears;
  deduplicate and count.

Output Format
Output ONLY the CSV with a header row and no explanation.
source,target,label,weight

Example
Input:  "The police cleared protesters from the bridge after the demonstration."
Output: source,target,label,weight
        police,clear,s,1
        clear,protester,o,1
        clear,bridge,from,1
        clear,demonstration,after,1
\end{sepprompt}
\section{Extractor Configuration}
\label{app:extractors}
We run SEP with three language models served through a common gateway, so only
the model backend changes and any performance difference reflects the model
rather than the parsing code: Haiku (\texttt{aws/claude4\_5\_haiku},
\texttt{claude-haiku-4-5-20251001}), GPT-4.1 (\texttt{openai/gpt4\_1}), and
Gemini (\texttt{gcp-deepmind/geminipro3\_1}). All three are called with
temperature $0.0$ for low-variance extraction, with a single call per article.
Coreference and normalization are not separate stages: the prompt of
Appendix~\ref{app:prompt} instructs the model to replace pronouns with their
referent, resolve short names to their longest form, and normalize tokens
(lowercasing, infinitive verbs, singular nouns, multi-word entities as single
nodes) within the same pass that emits the edge list. As a worked example, the
sentence ``The police cleared protesters from the bridge after the
demonstration'' yields the edges \texttt{police}$\to$\texttt{clear}~(s),
\texttt{clear}$\to$\texttt{protester}~(o),
\texttt{clear}$\to$\texttt{bridge}~(from), and
\texttt{clear}$\to$\texttt{demonstration}~(after).

\section{Per-Extractor Results}
\label{app:perextractor}
Tables~\ref{tab:perext-convoy}--\ref{tab:perext-melissa} report ROC-AUC / PR-AUC
separately for each extractor on the emergent protocol ($W=14$), supporting
the claim in Section~\ref{sec:robustness} that the scorer ranking does not
depend on the extractor. Preferential Attachment is the top scorer under every
extractor on every event; the three-extractor means are Table~\ref{tab:emergent}.

\begin{table}[t]\centering\footnotesize
\caption{Per-extractor ROC-AUC / PR-AUC, Freedom Convoy (emergent, $W=14$).}
\setlength{\tabcolsep}{4pt}
\begin{tabular}{lccc}\toprule
Scorer & Haiku & GPT-4.1 & Gemini\\ \midrule
Common Neighbors        & 0.728/0.378 & 0.757/0.426 & 0.726/0.382\\
Adamic--Adar            & 0.730/0.394 & 0.760/0.446 & 0.729/0.404\\
Resource Allocation     & 0.727/0.384 & 0.758/0.436 & 0.727/0.393\\
Jaccard                 & 0.678/0.290 & 0.708/0.317 & 0.675/0.275\\
Hub-Depressed           & 0.663/0.272 & 0.693/0.297 & 0.668/0.266\\
Preferential Attachment & \textbf{0.764/0.424} & \textbf{0.796/0.473} & \textbf{0.795/0.470}\\
RWR (Corman)            & 0.689/0.318 & 0.707/0.332 & 0.712/0.338\\
EdgeBank                & 0.637/0.299 & 0.639/0.311 & 0.642/0.329\\
\bottomrule
\end{tabular}

\label{tab:perext-convoy}
\end{table}

\begin{table}[t]\centering\footnotesize
\caption{Per-extractor ROC-AUC / PR-AUC, Iranian protests (emergent, $W=14$).}
\setlength{\tabcolsep}{4pt}
\begin{tabular}{lccc}\toprule
Scorer & Haiku & GPT-4.1 & Gemini\\ \midrule
Common Neighbors        & 0.710/0.344 & 0.745/0.391 & 0.709/0.352\\
Adamic--Adar            & 0.711/0.353 & 0.747/0.406 & 0.711/0.372\\
Resource Allocation     & 0.707/0.348 & 0.742/0.398 & 0.707/0.359\\
Jaccard                 & 0.662/0.285 & 0.700/0.316 & 0.658/0.255\\
Hub-Depressed           & 0.644/0.267 & 0.685/0.294 & 0.653/0.247\\
Preferential Attachment & \textbf{0.730/0.370} & \textbf{0.764/0.416} & \textbf{0.771/0.428}\\
RWR (Corman)            & 0.664/0.289 & 0.685/0.304 & 0.694/0.315\\
EdgeBank                & 0.645/0.284 & 0.643/0.295 & 0.638/0.315\\
\bottomrule
\end{tabular}

\label{tab:perext-iran}
\end{table}

\begin{table}[t]\centering\footnotesize
\caption{Per-extractor ROC-AUC / PR-AUC, Hurricane Melissa (emergent, $W=14$).}
\setlength{\tabcolsep}{4pt}
\begin{tabular}{lccc}\toprule
Scorer & Haiku & GPT-4.1 & Gemini\\ \midrule
Common Neighbors        & 0.694/0.352 & 0.718/0.401 & 0.655/0.321\\
Adamic--Adar            & 0.694/0.375 & 0.720/0.427 & 0.655/0.339\\
Resource Allocation     & 0.688/0.352 & 0.716/0.400 & 0.653/0.322\\
Jaccard                 & 0.653/0.246 & 0.685/0.274 & 0.639/0.238\\
Hub-Depressed           & 0.650/0.239 & 0.683/0.269 & 0.639/0.241\\
Preferential Attachment & \textbf{0.767/0.440} & \textbf{0.804/0.499} & \textbf{0.808/0.515}\\
RWR (Corman)            & 0.702/0.340 & 0.730/0.369 & 0.749/0.394\\
EdgeBank                & 0.681/0.350 & 0.671/0.355 & 0.687/0.399\\
\bottomrule
\end{tabular}

\label{tab:perext-melissa}
\end{table}

\section{Precision@$k$}
\label{app:preck}
Candidates with equal scores are ranked in uniformly random order (fixed seed).
This is immaterial for the structural indices, whose scores are effectively
continuous, but it matters for near-binary predictors such as EdgeBank, which
assigns the same score to most candidates.
\begin{table}[H]\centering\small
\caption{Precision@$k$ ($k$ = number of positives), mean over the three extractors, emergent protocol, $W=14$. Preferential Attachment is top on every event, consistent with the ROC-AUC / PR-AUC ranking.}
\setlength{\tabcolsep}{6pt}
\begin{tabular}{lccc}\toprule
Scorer & Convoy & Iran & Melissa\\ \midrule
Common Neighbors        & 0.429 & 0.397 & 0.414\\
Adamic--Adar            & 0.432 & 0.398 & 0.415\\
Resource Allocation     & 0.426 & 0.388 & 0.403\\
Jaccard                 & 0.335 & 0.311 & 0.326\\
Hub-Depressed           & 0.316 & 0.292 & 0.318\\
Preferential Attachment & \textbf{0.457} & \textbf{0.417} & \textbf{0.482}\\
RWR (Corman)            & 0.351 & 0.328 & 0.385\\
EdgeBank (memorization) & 0.384 & 0.389 & 0.441\\
\bottomrule
\end{tabular}

\label{tab:precisionk}
\end{table}

\section{Weighted vs.\ Unweighted Indices}
\label{app:weighted}
\begin{table}[H]\centering\small
\setlength{\tabcolsep}{5pt}
\begin{tabular}{lccc}\toprule
Scorer & Convoy & Iran & Melissa\\ \midrule
Common Neighbors        & 0.737 / 0.727 & 0.721 / 0.710 & 0.689 / 0.688\\
Adamic--Adar            & 0.740 / 0.730 & 0.723 / 0.712 & 0.690 / 0.689\\
Preferential Attachment & 0.785 / 0.751 & 0.755 / 0.723 & 0.793 / 0.778\\
RWR (Corman)            & 0.703 / 0.687 & 0.681 / 0.665 & 0.727 / 0.719\\
\bottomrule
\end{tabular}
\caption{Weighted vs.\ unweighted scorers (ROC-AUC, mean over three extractors,
emergent protocol). Each cell is unweighted / weighted. Recurrence weighting
never helps and slightly hurts, most for Preferential Attachment.}
\label{tab:weighted}
\end{table}

\section{Data Collection}
\label{app:data}
All three corpora were collected from the GDELT~2.0 Global Knowledge Graph
(GKG) via the BigQuery table
\path{gdelt-bq.gdeltv2.gkg_partitioned}. For each
event we exported a date-stamped URL list with an event-specific filter,
retrieved the full body text of each URL with trafilatura, and removed
near-duplicate articles by deduplicating on the normalized article text, which
collapses syndicated wire copy. Every query applies no per-day row cap, so the
exported list is the full census over the window. The Convoy and Melissa queries
key off the URL slug and restrict to English-origin documents
(\texttt{TranslationInfo IS NULL}). The Iran query keys off language-independent GKG themes and locations and does not apply this filter; nonetheless, the retrieved full-text corpus is over 99\% English, so all three events are English-language. Realized corpus sizes are in
Table~\ref{tab:corpus}. The Iran window's upper bound was the export date; see
Table~\ref{tab:corpus} for its realized day count.

\medskip\noindent{\bfseries Freedom Convoy (2022-01-20 to 2022-04-01)}\par\nobreak\smallskip
\begin{lstlisting}[style=sqlpretty]
WITH art AS (
  SELECT
    FORMAT_DATE('%Y-%m-%d',
      PARSE_DATE('%Y%m%d', SUBSTR(CAST(DATE AS STRING),1,8))) AS date,
    DocumentIdentifier          AS url,
    ANY_VALUE(SourceCommonName) AS domain,
    COUNT(*)                    AS mentions
  FROM `gdelt-bq.gdeltv2.gkg_partitioned`
  WHERE _PARTITIONTIME BETWEEN TIMESTAMP('2022-01-20')
                           AND TIMESTAMP('2022-04-01')
    AND DocumentIdentifier LIKE 'http%'
    AND (TranslationInfo IS NULL OR TranslationInfo = '')
    AND LOWER(V2Locations) LIKE '%canada%'
    AND ( LOWER(DocumentIdentifier) LIKE '%convoy%'
       OR LOWER(DocumentIdentifier) LIKE '%trucker%'
       OR LOWER(V2Organizations)  LIKE '%freedom convoy%' )
  GROUP BY date, url
)
SELECT date, url, domain FROM art ORDER BY date, mentions DESC;
\end{lstlisting}

\medskip\noindent{\bfseries Iranian protests (2025-12-25 onward)}\par\nobreak\smallskip
\begin{lstlisting}[style=sqlpretty]
WITH art AS (
  SELECT
    FORMAT_DATE('%Y-%m-%d',
      PARSE_DATE('%Y%m%d', SUBSTR(CAST(DATE AS STRING),1,8))) AS date,
    DocumentIdentifier          AS url,
    ANY_VALUE(SourceCommonName) AS domain,
    COUNT(*)                    AS mentions
  FROM `gdelt-bq.gdeltv2.gkg_partitioned`
  WHERE _PARTITIONTIME BETWEEN TIMESTAMP('2025-12-25')
                           AND CURRENT_TIMESTAMP()
    AND DocumentIdentifier LIKE 'http%'
    AND (LOWER(V2Locations) LIKE '%iran%'
         OR V2Locations LIKE '%#IR#%')
    AND ( V2Themes LIKE '%PROTEST%'
       OR V2Themes LIKE '%UNREST%'
       OR V2Themes LIKE '%CRACKDOWN%' )
  GROUP BY date, url
)
SELECT date, url, domain FROM art ORDER BY date, mentions DESC;
\end{lstlisting}

\medskip\noindent{\bfseries Hurricane Melissa (2025-10-25 to 2025-12-08)}\par\nobreak\smallskip
\begin{lstlisting}[style=sqlpretty]
WITH art AS (
  SELECT
    FORMAT_DATE('%Y-%m-%d',
      PARSE_DATE('%Y%m%d', SUBSTR(CAST(DATE AS STRING),1,8))) AS date,
    DocumentIdentifier          AS url,
    ANY_VALUE(SourceCommonName) AS domain,
    COUNT(*)                    AS mentions
  FROM `gdelt-bq.gdeltv2.gkg_partitioned`
  WHERE _PARTITIONTIME BETWEEN TIMESTAMP('2025-10-25')
                           AND TIMESTAMP('2025-12-08')
    AND DocumentIdentifier LIKE 'http%'
    AND (TranslationInfo IS NULL OR TranslationInfo = '')
    AND LOWER(DocumentIdentifier) LIKE '%melissa%'
    AND ( LOWER(DocumentIdentifier) LIKE '%hurricane%'
       OR LOWER(DocumentIdentifier) LIKE '%storm%'
       OR LOWER(V2Locations)        LIKE '%jamaica%' )
  GROUP BY date, url
)
SELECT date, url, domain FROM art ORDER BY date, mentions DESC;
\end{lstlisting}

\section{Baselines and Hyperparameters}
\label{app:hyper}
\paragraph{Structural scorers (emergent protocol).} All indices use a rolling
window $W=14$ and a $5:1$ negative ratio (positive base rate
$1/6\approx0.167$). The candidate set for each day is the unconnected ordered
pairs among the top $1{,}000$ nodes by combined in- and out-degree in the
window, after removing a stoplist of light verbs and generic nouns: \{be, say,
have, do, get, make, go, take, come, give, see, know, tell, want, person,
people, thing, time, year, day, way, more, one, two, week\}. Negatives are drawn
uniformly from unconnected ordered pairs. Days with fewer than $50$ nodes or
$10$ positive edges are skipped. All runs are fully deterministic: a fixed
random seed ($0$) together with a fixed candidate ordering, so the reported
numbers reproduce exactly.

\paragraph{Standard protocol.} EdgeBank is run in unlimited-memory mode: a
candidate is scored $1$ if the directed pair appeared anywhere in the training
window. TGN is run with DyGLib on the same top-$1{,}000$-node graph, using a
chronological $70/15/15$ train/validation/test split, $1{:}1$
destination-perturbed negatives, and DyGLib's default TGN training
configuration (\texttt{num\_runs}=1). Because TGN requires a single large
train/test split, it is evaluated only on the standard protocol, not the
day-by-day forward chaining used for the structural scorers.




\section{Rolling-Window Sensitivity: All Events}
\label{app:window}



\begin{figure}[H]\centering
  \includegraphics[width=0.5\textwidth]{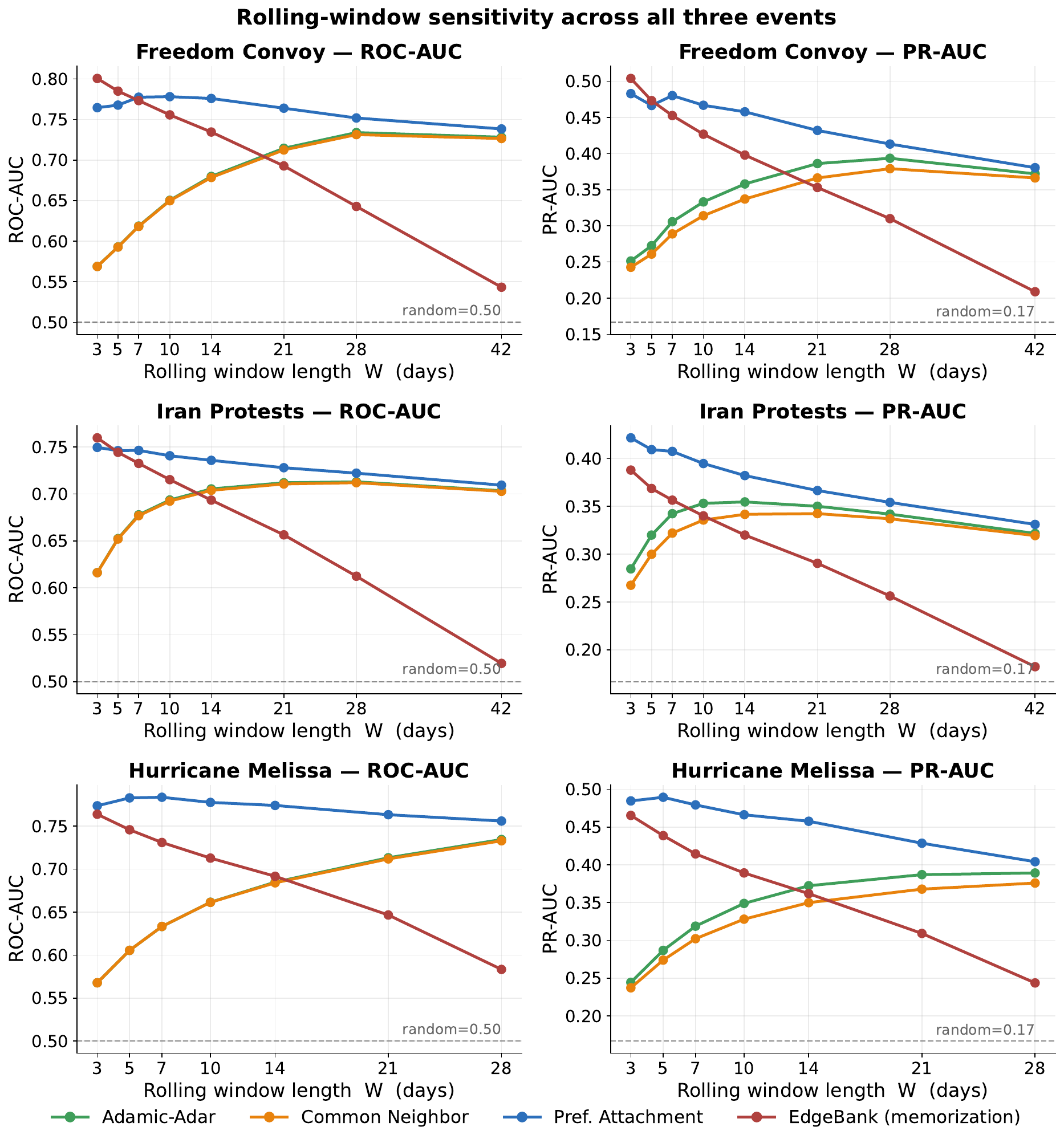}
    \caption{Rolling-window sensitivity for all three events (Haiku extractor;
  each evaluated day has at least $\max(W)$ days of history, so values are not
  directly comparable to the three-extractor means in Table~\ref{tab:emergent}).
  Across all three events, the advantage of structural scorers over EdgeBank
  grows as the window $W$ lengthens: EdgeBank decays toward chance while the
  structural scorers stay flat or improve.}
  \label{fig:window_all}
\end{figure}

Fig.~\ref{fig:window_all} sweeps the rolling-window length $W$ under the same
protocol. As the window grows, the EdgeBank memorization baseline declines
steeply toward chance, because a longer window absorbs more of the recurring
edges and leaves proportionally fewer of them to reappear at $t+1$. The
structural scorers stay comparatively flat, and Preferential Attachment even
rises on Hurricane Melissa. The margin between structure and memorization
therefore widens with $W$, and the pattern is consistent across the three
events.

\end{document}